\documentclass[12pt,preprint]{aastex}
\begin{document}

\title{Hyperon bulk viscosity in the presence of antikaon condensate}
\author{Debarati Chatterjee\altaffilmark{1} and 
Debades Bandyopadhyay\altaffilmark{1}} 
\affil{Theory Division, Saha Institute of Nuclear Physics, 1/AF Bidhannagar, 
Kolkata-700064, India}
\altaffiltext{1}{Centre for Astroparticle Physics, Saha Institute of Nuclear
Physics, 1/AF Bidhannagar, Kolkata-700064, India}

\begin{abstract}
We investigate the hyperon bulk viscosity due to the non-leptonic process
$n + p \rightleftharpoons p + \Lambda~$ in $K^-$ condensed matter and its 
effect on the r-mode instability in neutron stars. We find that the hyperon 
bulk viscosity coefficient in the presence of antikaon condensate is suppressed 
compared with the case without the condensate. The suppressed 
hyperon bulk viscosity in the superconducting phase is still an efficient 
mechanism to damp the r-mode instability in neutron stars. 
\end{abstract}


\section{Introduction}

The study of r-modes in rotating neutron stars has generated great interest
in recent times from two points of view. On one hand, r-modes of rotating 
neutron stars could be possible sources of detectable gravitational waves. On 
the other hand, the gravitational radiation driven instability of r-modes 
\citep{Chan,Fri1,Fri2,Fri3,Fri98,Lin98,And98,And99,Kok,And01,Ster} 
might play an important role in regulating spins of 
young as well as old, accreting neutron stars in low mass x-ray 
binaries(LMXBs). The latter situation has been strengthened by the fact that
the spin distribution of pulsars has an upper limit at 730 Hz \citep{Cha2,Cha}.
It is worth mentioning here that the fastest rotating pulsar has a spin 
frequency 716 Hz \citep{Hes}. 

The bulk viscosity of neutron star matter is an important issue in 
connection with the damping of the r-mode instability in rotating neutron stars.
The r-mode instability may be suppressed through different mechanisms. 
It was shown that the damping of the instability by the large hyperon bulk 
viscosity coefficient could be an efficient mechanism 
\citep{Jon1,Jon2,Lin02,Han3,Dal,Dra,Nar,DR1,DR2}. However, it may be suppressed
by several orders of magnitude when neutrons, protons or
hyperons are superfluid \citep{Han1,Han2,Han3,Nar,And06}. This study was 
later extended to the
calculation of bulk viscosity due to other exotic forms of matter
such as unpaired quark matter \citep{Mad92,Mad00,Don1,Don2}, quark-hadron 
mixed phase \citep{Dra,Pan} and color superconducting quark matter 
\citep{Schm,Alf,Alf3,Bas,Bas2}. Another 
possibility of damping the r-mode instability is the mutual friction between
inter-penetrating neutron and proton superfluids \citep{Lin00,And06}. 

The superfluidity of neutron star matter plays a significant role on the
damping of the r-modes in neutron stars. It was shown that superfluid particles
taking part in non-leptonic weak processes involving hyperons might reduce the 
hyperon bulk viscosity coefficient by several orders of magnitude 
\citep{Han3,And06}.  
Recently, we have investigated the bulk viscosity due to the non-leptonic
weak process $n \rightleftharpoons p + K^{-}$ in $K^-$ condensed matter
\citep{DR3}. 
The kaon bulk viscosity was suppressed in the condensed phase by several orders
of magnitude and could not damp the r-mode instability. This motivates us 
to further investigate how the bulk viscosity due to the 
non-leptonic process $n + p \rightleftharpoons p + \Lambda~$ behaves in 
antikaon condensed matter and how it influences r-modes of neutron stars.  

We organise the paper in the following way. In section 2, the 
field theoretical models of strong interactions, different phases of dense 
matter, the bulk viscosity coefficient and the corresponding time scale are
described. Results of our calculation are discussed in section 3. Section 4 
gives summary and conclusions. 

\section{Formalism}
Our motivation is to calculate the hyperon bulk viscosity due to the 
non-leptonic process $n + p \rightleftharpoons p + \Lambda~$ in a 
superconducting phase i.e. antikaon condensed phase. This process was 
extensively investigated in hadronic matter by several groups 
\citep{Jon1,Jon2,Lin02,Dal,Nar,DR1,DR2}. In this case,
we consider a first order phase transition from hadronic 
to $K^-$ condensed matter. Both 
the pure hadronic and $K^-$ condensed matter are described within the
framework of relativistic field theoretical models. The constituents 
of matter in both phases are neutrons ($n$), protons ($p$), $\Lambda$ hyperons,
electrons and muons. The baryon-baryon interaction is mediated by the exchange
of $\sigma$, $\omega$ and $\rho$ mesons and two strange mesons, scalar meson, 
$f_0$(975) (denoted hereafter as $\sigma^*$) and the vector meson 
$\phi$(1020) \citep{Sch}. 
Both phases maintain  local charge neutrality and  beta-equilibrium conditions 
whereas these conditions are satisfied globally in the mixed phase. Further 
baryons are
embedded in the $K^-$ condensate phase. We describe the equation of state (EoS)
in the antikaon condensed phase using the following Lagrangian density for 
baryons \citep{Sch}
\begin{eqnarray}
{\cal L}_B &=& \sum_B \bar\Psi_{B}\left(i\gamma_\mu{\partial^\mu} - m_B
+ g_{\sigma B} \sigma - g_{\omega B} \gamma_\mu \omega^\mu
- g_{\rho B}
\gamma_\mu{\mbox{\boldmath t}}_B \cdot
{\mbox{\boldmath $\rho$}}^\mu \right)\Psi_B\nonumber\\
&& + \frac{1}{2}\left( \partial_\mu \sigma\partial^\mu \sigma
- m_\sigma^2 \sigma^2\right) - U(\sigma) \nonumber\\
&& -\frac{1}{4} \omega_{\mu\nu}\omega^{\mu\nu}
+\frac{1}{2}m_\omega^2 \omega_\mu \omega^\mu
- \frac{1}{4}{\mbox {\boldmath $\rho$}}_{\mu\nu} \cdot
{\mbox {\boldmath $\rho$}}^{\mu\nu}
+ \frac{1}{2}m_\rho^2 {\mbox {\boldmath $\rho$}}_\mu \cdot
{\mbox {\boldmath $\rho$}}^\mu + {\cal L}_{YY}~.
\end{eqnarray}
Here $\psi_B$ denotes the Dirac bispinor for baryons $B$ with vacuum mass $m_B$
and the isospin operator is ${\mbox {\boldmath t}}_B$. The scalar
self-interaction term \citep{Bog} is
\begin{equation}
U(\sigma) = \frac{1}{3} g_2 \sigma^3 + \frac{1}{4} g_3 \sigma^4 ~.
\end{equation}
The Lagrangian density for hyperon-hyperon interaction (${\cal L}_{YY}$)
is given by
\begin{eqnarray}
{\cal L}_{YY} &=& \sum_B \bar\Psi_{B}\left(
g_{\sigma^* B} \sigma^* - g_{\phi B} \gamma_\mu \phi^\mu
\right)\Psi_B\nonumber\\
&& + \frac{1}{2}\left( \partial_\mu \sigma^*\partial^\mu \sigma^*
- m_{\sigma^*}^2 \sigma^{*2}\right)
-\frac{1}{4} \phi_{\mu\nu}\phi^{\mu\nu}
+\frac{1}{2}m_\phi^2 \phi_\mu \phi^\mu~.
\end{eqnarray}
As nucleons do not couple with strange mesons, 
$g_{\sigma^* B} = g_{\phi B} = 0$.

Similarly we treat the (anti)kaon-baryon interaction in the same footing as the 
baryon-baryon interaction. The Lagrangian density for (anti)kaons in the 
minimal coupling scheme is \citep{Gle98,Gle99,Bani1,Bani2}, 
\begin{equation}
{\cal L}_K = D^*_\mu{\bar K} D^\mu K - m_K^{* 2} {\bar K} K ~,
\end{equation}
where the covariant derivative is
$D_\mu = \partial_\mu + ig_{\omega K}{\omega_\mu} + ig_{\phi K}{\phi_\mu}
+ i g_{\rho K}
{\mbox{\boldmath t}}_K \cdot {\mbox{\boldmath $\rho$}}_\mu$ and
the effective mass of (anti)kaons is 
$m_K^* = m_K - g_{\sigma K} \sigma - g_{\sigma^* K} \sigma^*$.

We perform this calculation in the mean field approximation 
\citep{Ser}. 
The mean meson fields in the condensed phase are denoted by $\sigma$, 
$\sigma^*$, $\omega_0$, $\phi_0$ and $\rho_{03}$. The expressions for mean 
fields can be found in Ref. \citep{Bani2,DR3}. 
The in-medium energy of
$K^-$ mesons for $s$-wave (${\vec k}=0$) condensation is given by
\begin{equation}
\omega_{K^-} = \mu_{K^-} = m_K^* - g_{\omega K} \omega_0 - g_{\phi K} \phi_0
+ I_{3K^-} g_{\rho K} \rho_{03} ~,
\end{equation}
where $\mu_{K^-}$ is the chemical potential of $K^-$ mesons and the isospin 
projection is $I_{3K^-} = -1/2$. 
The chemical potential for baryons $B$ is given by
\begin{equation}
\mu_{B}^{K} = \left(k^2_{F_{B}} + {m_B^{*K}}^2 \right)^{1/2} + g_{\omega B} 
\omega_0 + g_{\phi B} \phi_0 + I_{3B} g_{\rho B} \rho_{03}~
\end{equation}
where effective baryon mass is $m_B^{*K}=m_B - g_{\sigma B}\sigma 
- g_{\sigma^* B} \sigma^*$ and
isospin projection for baryons $B$ is $I_{3B}$. 

We obtain the mean fields in the hadronic phase putting 
source terms for $K^-$ mesons equal to zero in corresponding equations of 
motion \citep{Bani2,DR3}. 

The total energy density and pressure in the antikaon condensed phase are given
by 
\begin{eqnarray}
{\varepsilon^{K}}  &=& \frac{1}{2}m_\sigma^2 \sigma^2
+ \frac{1}{3} g_2 \sigma^3
+ \frac{1}{4} g_3 \sigma^4  + \frac{1}{2}m_{\sigma^*}^2 \sigma^{*2}
+ \frac{1}{2} m_\omega^2 \omega_0^2 + \frac{1}{2} m_\phi^2 \phi_0^2
+ \frac{1}{2} m_\rho^2 \rho_{03}^2  \nonumber \\
&& + \sum_{B=n,p,\Lambda} \frac{2J_B+1}{2\pi^2}
\int_0^{k_{F_B}} (k^2+{m_B^{*K}}^2)^{1/2} k^2 \ dk
+ \sum_{l=e^-,\mu^-} \frac{1}{\pi^2} \int_0^{K_{F_l}} (k^2+m^2_l)^{1/2} k^2 \ dk
\nonumber \\
 && + m^*_K n_{K^-}~,
\end{eqnarray}
and
\begin{eqnarray}
P^{K} &=& - \frac{1}{2}m_\sigma^2 \sigma^2 - \frac{1}{3} g_2 \sigma^3
- \frac{1}{4} g_3 \sigma^4  - \frac{1}{2}m_{\sigma^*}^2 \sigma^{*2}
+ \frac{1}{2} m_\omega^2 \omega_0^2 + \frac{1}{2} m_\phi^2 \phi_0^2
+ \frac{1}{2} m_\rho^2 \rho_{03}^2 \nonumber \\
&& + \frac{1}{3}\sum_{B=n,p,\Lambda} \frac{2J_B+1}{2\pi^2}
\int_0^{k_{F_B}} \frac{k^4 \ dk}{(k^2+{m_B^{*K}}^ 2)^{1/2}}
+ \frac{1}{3} \sum_{l=e^-,\mu^-} \frac{1}{\pi^2}
\int_0^{K_{F_l}} \frac{k^4 \ dk}{(k^2+m^2_l)^{1/2}}~.
\end{eqnarray}

We describe the mixed phase of hadronic and $K^-$ condensed matter using the 
Gibbs conditions for thermodynamic equilibrium along with global charge and
baryon number conservation laws 
\citep{Gle92,Gle99}. 

Now we focus on the calculation of bulk viscosity due to the non-leptonic
process. It was shown that the real part of bulk viscosity coefficient was 
related to relaxation times of microscopic processes \citep{Lan,Lin02} by the
relation, 
\begin{equation}
\zeta = \frac {P(\gamma_{\infty} - \gamma_0)\tau}{1 + {(\omega\tau)}^2}~,
\end{equation}
where $P$ is the pressure, $\tau$ is the net microscopic relaxation time and 
$\gamma_{\infty}$ and $\gamma_0$ are 'infinite' and 'zero' frequency adiabatic 
indices respectively and their difference 
\begin{equation}
\gamma_{\infty} - \gamma_0 = - \frac {n_b^2}{P} \frac {\partial P} 
{\partial n_n} \frac {d{\bar x}_n} {dn_b}~,
\end{equation}
is determined using EoS as input. Here $\bar x_n = \frac {n_n}{n_b}$ gives the 
neutron fraction in the equilibrium state and $n_b = {\sum}_{B} n_B$ is the 
total baryon density. We are interested in (l,m) r-mode. In this case, 
the angular velocity ($\omega$) of the mode is related to the angular velocity 
($\Omega$) of a rotating neutron star as 
$\omega = {\frac {2m}{l(l+1)}} \Omega$ \citep{And01}. 

The partial derivatives of pressure with respect to neutron fraction in both 
phases are calculated using the Gibbs-Duhem relation \citep{DR3}. 
In the pure hadronic phase, this relation gives
\begin{equation}
\frac{\partial P^h}{\partial n_n^h} = n_n^h \alpha_{nn}^h 
+  n_p^h \alpha_{pn}^h + n_{\Lambda}^h \alpha_{\Lambda n}^h~,
\end{equation} 
where $\alpha_{ij}$ is defined by
$\alpha_{ij} = \left(\frac{\partial \mu_i}{\partial n_j}\right)_{n_k,k\neq j}$. 
Similarly, in the pure condensed phase, it is given by 
\begin{equation}
\frac{\partial P^K}{\partial n_n^K} = n_n^K \alpha_{nn}^K 
+  n_p^K \alpha_{pn}^K 
+ n_{\Lambda}^K \alpha_{\Lambda n}^K
+ n_{K^-} \alpha_{K^- n}^K~.
\end{equation} 
In the mixed phase, this relation has the form
\begin{equation}
\frac{\partial P}{\partial n_n} = 
\frac{\partial P^h}{\partial n_n^h} +  
\frac{\partial P^K}{\partial n_n^K}~, 
\end{equation} 
where $n_n = (1 - \chi) n_n^h + \chi n_n^K$.

Next we calculate relaxation times of the non-leptonic process in different 
phases. In this case, we express all perturbed quantities
in terms of the variation in neutron number density ($n_n$) in the 
respective phase. The relaxation time ($\tau$) for the 
non-leptonic process is given by \citep{Lin02}
\begin{equation}
\frac{1}{\tau} = \frac{\Gamma_{\Lambda}}{\delta \mu} 
\frac{\delta \mu}{\delta n_n^j}~.
\end{equation}
Here, $\delta {n_n^j} = n_n^j - {\bar n}_n^j$ is the departure of neutron 
fraction from its thermodynamic equilibrium value ${\bar n}_n^j$ in the
j-th (=hadron, $K^-$ condensed) phase. The reaction rate per unit volume for 
the non-leptonic process in question was already calculated by others 
\citep{Lin02,Nar}. The relaxation time ($\tau$) for the process in the hadronic 
phase is given by \citep{Lin02}
\begin{equation}
\frac {1}{\tau} = \frac {{(kT)}^2}{192{\pi}^3} {p_{\Lambda}}
{<{{|M_{\Lambda}|}^2}>} \frac {\delta \mu}{\delta{n_n^h}}~,
\end{equation}
along with
\begin{equation}
\frac {\delta \mu}{\delta{n_n^h}} = 
\left(\alpha_{nn}^h - \alpha_{\Lambda n}^h\right)
+ \left(\alpha_{np}^h - \alpha_{\Lambda p}^h\right)
\frac {\delta {n_p^h}}{\delta{n_n^h}} 
+ \left(\alpha_{n \Lambda}^h - \alpha_{\Lambda \Lambda}^h\right)
\frac {\delta {n_{\Lambda}^h}}{\delta{n_n^h}}~,
\end{equation}
where $p_{\Lambda}$ is the Fermi momentum for $\Lambda$ hyperons and 
$<{|M_{\Lambda}|}^2>$ is the angle averaged matrix element squared in the
hadronic phase given by Ref.\citep{Lin02,Nar}. In pure hadronic phase, the 
second term in Eq.(16) vanishes because $\delta n_p^h = 0$. However, the 
calculation of the hadronic part of the mixed phase is a little bit involved
and described below.   

The relaxation time in the antikaon condensed phase has the same 
form as in Eq.(15). In this case,
the angle averaged matrix element , $<{|M_{\Lambda}|}^2>$ and $p_{\Lambda}$ are
to be calculated in the condensed phase. Also, we calculate
$\frac{\delta \mu}{\delta {n_n^K}}$ from the chemical potential imbalance due 
to the non-leptonic hyperon process $n + p \rightleftharpoons p + \Lambda~$ and
it is given by,
\begin{eqnarray}
\delta \mu &=& \delta \mu_n^K - \delta \mu_{\Lambda}^K \nonumber \\
&=& \left( \alpha_{nn}^K \delta n_n^K + \alpha_{np}^K \delta n_p^K  
+ \alpha_{n \Lambda}^K \delta n_{\Lambda}^K + \alpha_{n K^-} \delta n_{K^-}
\right) \nonumber\\ 
&-& \left( \alpha_{\Lambda n}^K \delta n_n^K + \alpha_{\Lambda p}^K 
\delta n_p^K  + \alpha_{\Lambda \Lambda}^K \delta n_{\Lambda}^K 
+ \alpha_{n K^-} \delta n_{K^-}\right)~.
\end{eqnarray}
We express $\delta {\mu}$ in terms of $\delta{n_n^K}$ and obtain 
$\frac {\delta \mu}{\delta{n_n^K}}$ using the following constraints,
\begin{eqnarray}
\left(\delta n_n^K + \delta n_p^K + \delta n_{\Lambda}^K \right) 
&=& 0~, \nonumber \\
\left(\delta n_p^K - \delta n_{K^-}\right) &=& 0~. 
\end{eqnarray}
and the chemical equilibrium in the strangeness changing process
$n \rightleftharpoons p + K^-$, 
\begin{eqnarray}
\delta \mu_n^K - \delta \mu_p^K - \delta \mu_{K^-} 
&=& \left( \alpha_{nn}^K \delta n_n^K + \alpha_{np}^K \delta n_p^K + 
+ \alpha_{n \Lambda}^K \delta n_{\Lambda}^K 
+ \alpha_{n K^-} \delta n_{K^-} \right) \nonumber \\
&-& \left( \alpha_{pn}^K \delta n_n^K + \alpha_{pp}^K \delta n_p^K + 
+ \alpha_{p \Lambda}^K \delta n_{\Lambda}^K 
+ \alpha_{p K^-} \delta n_{K^-} \right) \nonumber \\
&-& \left( \alpha_{K^- n} \delta n_n^K + \alpha_{K^- p} \delta n_p^K + 
+ \alpha_{K^- \Lambda} \delta n_{\Lambda}^K 
+ \alpha_{K^- K^-} \delta n_{K^-} \right) = 0~.
\end{eqnarray}

Next we calculate $\alpha_{ij}$ in the hadronic as well as $K^-$ condensed
phases using the EoS. We can write down these quantities for both phases in 
generalised forms. For $B = B'$, we get
\begin{eqnarray}
\alpha_{BB'}^P &=& \frac{\partial \mu_B^P}{\partial n_{B'}^P} \nonumber \\
&=& \left(\frac{g_{\omega B}}{m_{\omega}}\right)^2 
+ \frac{1}{4}\left(\frac{g_{\rho B}}{m_{\rho}}\right)^2 
+ \left(\frac{g_{\phi B}}{m_{\phi}}\right)^2 
+ \frac{\pi^2}{k_{F_B} \sqrt{k_{F_B}^2 
+ {m_B^{*P}}^2}} \nonumber \\ 
&-& {\frac {m_B^{*P}}{\sqrt{k_{F_B}^2 + 
{m_B^{*P}}^2}}}\left(g_{\sigma B}{\frac{\partial \sigma}{\partial n_{B'}^P}}
+ g_{\sigma^* B} {\frac {\partial \sigma^*}{\partial n_{B'}^P}}\right)~,
\end{eqnarray} 
and for $B \neq B'$
\begin{eqnarray}
\alpha_{BB'}^P &=& \frac{\partial \mu_B^P}{\partial n_{B'}^P} \nonumber \\
&=& \left(\frac{g_{\omega B} g_{\omega B'}}{m_{\omega}^2}\right) 
- \frac{1}{4}\left(\frac{g_{\rho B} g_{\rho B'}}{m_{\rho}^2}\right)\nonumber \\ 
&-& {\frac {m_B^{*P}}{\sqrt{k_{F_B}^2 + 
{m_B^{*P}}^2}}}\left(g_{\sigma B}{\frac{\partial \sigma}{\partial n_{B'}^P}}
+ g_{\sigma^* B} {\frac {\partial \sigma^*}{\partial n_{B'}^P}}\right)~,
\end{eqnarray} 
along with the following relations applicable for both cases
\begin{equation}
\frac{\partial \sigma}{\partial n_{B'}^P} = 
\frac{\left(\frac{g_{\sigma B}}{m_{\sigma}^2}\right)
\frac{m_B^{*P}}{\sqrt{k_{F_B}^2 + {m_B^{*P}}^2}}}{D - D' \times D''}~,
\end{equation}
\begin{equation}
\frac{\partial \sigma^*}{\partial n_{B'}^P} = -D''  
\frac{\partial \sigma}{\partial n_{B'}^P}~,  
\end{equation}
where
\begin{equation}
D = { 1+\frac{1}{m_\sigma^2}\frac{d^2U}{d \sigma^2} 
+ \sum_{B=n,p,\Lambda} \frac{(2J_B+1)}{2 \pi^2}\left(\frac{g_{\sigma B}}
{m_\sigma}
\right)^2 \int_0^{K_{F_B}} \frac{k^4 dk}{(k^2 + {m_B^{*P}}^2)^{3/2}}}~,
\end{equation} 
\begin{equation}
D' = \sum_{B=n,p,\Lambda} \frac{(2J_B+1)}{2 \pi^2}
\left(\frac{g_{\sigma B}g_{\sigma^* B}}{{m_\sigma}^2}
\right) \int_0^{K_{F_B}} \frac{k^4 dk}{(k^2 + {m_B^{*P}}^2)^{3/2}}~,
\end{equation} 
\begin{equation}
D'' = \frac {\sum_{B=n,p,\Lambda} \frac{(2J_B+1)}{2 \pi^2}
\left(\frac{g_{\sigma B}g_{\sigma^* B}}{{m_\sigma^*}^2}
\right) \int_0^{K_{F_B}} \frac{k^4 dk}{(k^2 + {m_B^{*P}}^2)^{3/2}}}
{ 1+ \sum_{B=n,p,\Lambda} \frac{(2J_B+1)}{2 \pi^2}\left(\frac{g_{\sigma^* B}}
{m_\sigma^*}
\right)^2 \int_0^{K_{F_B}} \frac{k^4 dk}{(k^2 + {m_B^{*P}}^2)^{3/2}}}~.
\end{equation} 
Here $B$ and $B'$ denote baryons and P stands for hadron ($h$) or antikaon ($K$)
phase. Further nucleons do not couple with strange strange mesons i.e.
$g_{\sigma^* N} = g_{\phi N} = 0$. Similarly, $\Lambda$ hyperons do not couple 
with $\rho$ meson i.e. $g_{\rho \Lambda} = 0$. The results for other $\alpha$s
in the antikaon condensed phase are given below,
\begin{eqnarray}
\alpha_{B K^-} &=& \frac{\partial \mu_B^K}{\partial n_{K^-}} \nonumber \\
&=& -\left( \frac{g_{\omega B}g_{\omega K}}{m_\omega^2}\right) 
\pm \frac{1}{4}\left( \frac{g_{\rho B}g_{\rho K}}{m_\rho^2}\right) 
-\left( \frac{g_{\phi B}g_{\phi K}}{m_\phi^2}\right) \nonumber \\ 
&-& {\frac {m_B^{*K}}{\sqrt{k_{F_B}^2 + 
{m_B^{*K}}^2}}}\left(g_{\sigma B}{\frac{\partial \sigma}{\partial n_{K^-}}}
+ g_{\sigma^* B} {\frac {\partial \sigma^*}{\partial n_{K^-}}}\right)~,
\end{eqnarray}
\begin{eqnarray}
\alpha_{K^- B} &=& \frac{\partial \mu_{K^-}}{\partial n_B^K} \nonumber \\
&=& -\left( \frac{g_{\omega B}g_{\omega K}}{m_\omega^2}\right) 
\pm \frac{1}{4}\left( \frac{g_{\rho B}g_{\rho K}}{m_\rho^2}\right) 
-\left( \frac{g_{\phi B}g_{\phi K}}{m_\phi^2}\right) 
- \left(g_{\sigma K}{\frac{\partial \sigma}{\partial n_B^K}}
+ g_{\sigma^* K} {\frac {\partial \sigma^*}{\partial n_B^K}}\right)~,
\end{eqnarray}
\begin{equation}
\alpha_{K^- K^-} = \frac{\partial \mu_{K^-}}{\partial n_{K^-}} \\
= \left( \frac{g_{\omega K}}{m_{\omega}} \right)^2 
+ \left( \frac{g_{\phi K}}{m_{\phi}} \right)^2 
+ \frac{1}{4}\left( \frac{g_{\rho K}}{m_{\rho}} \right)^2 
- g_{\sigma K} \frac{\partial \sigma}{\partial n_{K^-}} 
- g_{\sigma^* K} \frac{\partial \sigma^{*}}{\partial n_{K^-}}~,
\end{equation}
where,
\begin{eqnarray}
\frac{\partial \sigma}{\partial n_{K^-}} &=& \frac{\frac{g_{\sigma K}}
{m_{\sigma}^2} - \frac{g_{\sigma^* K} D''}{m_{\sigma}^2}}{D- D' \times D''},
\nonumber\\
\frac{\partial \sigma^*}{\partial n_{K^-}} &=& \frac{\frac{g_{\sigma^* K}}
{m_{\sigma^*}^2}}{F} - D'' \frac{\partial \sigma}{\partial n_{K^-}}~, 
\end{eqnarray} 
and 
\begin{equation}
F = { 1 + \sum_{B=n,p,\Lambda} \frac{(2J_B+1)}{2 \pi^2}
\left(\frac{g_{\sigma^* B}} {m_\sigma^*}
\right)^2 \int_0^{K_{F_B}} \frac{k^4 dk}{(k^2 + {m_B^{*P}}^2)^{3/2}}}~.
\end{equation} 
In the second term in Eq.(27) and (28), +ve sign corresponds to neutrons and
-ve is for protons. With the given $\alpha_{ij}$, we can now calculate 
relaxation time for the non-leptonic process in hadron as well as antikaon 
condensed phases. As soon as we know the relaxation time, we can calculate the 
bulk viscosity coefficient in each phase. 

Now we focus on the calculation of relaxation time and bulk viscosity in the
mixed phase. For this, we have to express the chemical imbalance 
($\delta {\mu}$) in the non-leptonic hyperon process as given by Eq.(17) in 
terms of $\delta{n_n^K}$ from the following constraints,
\begin{eqnarray}
(1-\chi)\left(\delta n_n^h + \delta n_p^h + \delta n_{\Lambda}^h \right) 
+ \chi \left(\delta n_n^K + \delta n_p^K + \delta n_{\Lambda}^K \right) 
&=& 0~, \nonumber \\ 
(1-\chi)\delta n_p^h + \chi ( \delta n_p^K - \delta n_{K^-}) &=& 0~, 
\nonumber \\ 
\delta \mu_p^h &=& \delta \mu_p^K~, \nonumber \\ 
\delta \mu_n^h &=& \delta \mu_n^K~, \nonumber \\ 
\delta \mu_{\Lambda}^h &=& \delta \mu_{\Lambda}^K~, \nonumber \\ 
\delta \mu_n^K - \delta \mu_p^K - \delta \mu_{K^-} &=& 0~. 
\end{eqnarray} 
Here we have $\delta \chi = 0$ because number densities
deviate from their equilibrium values only by internal reactions \citep{Lin02}.
First two constraints follow from the conservation of baryon number and
electric charge neutrality. The last constrain is the result of the chemical
equilibrium involving $K^-$ condensate as already shown by Eq.(19). The 
other constraints are due to the equality of neutron, proton and $\Lambda$ 
chemical potentials in the hadronic and condensed phases and 
we can rewrite them as
\begin{eqnarray}
\left(\alpha_{pn}^h \delta n_n^h + \alpha_{pp}^h \delta n_p^h 
+ \alpha_{p\Lambda}^h \delta n_{\Lambda}^h\right) -   
\left(\alpha_{pn}^K \delta n_n^K + \alpha_{pp}^K \delta n_p^K 
+ \alpha_{p\Lambda}^K \delta n_{\Lambda}^K +   
\alpha_{p K^-} \delta n_{K^-}\right) &=& 0~, \nonumber \\ 
\left(\alpha_{nn}^h \delta n_n^h + \alpha_{np}^h \delta n_p^h 
+ \alpha_{n \Lambda}^h \delta n_{\Lambda}^h \right) -   
\left(\alpha_{nn}^K \delta n_n^K + \alpha_{np}^K \delta n_p^K 
+ \alpha_{n \Lambda}^K \delta n_{\Lambda}^K +    
\alpha_{n K^-} \delta n_{K^-} \right) &=& 0~, \nonumber \\ 
\left(\alpha_{\Lambda n}^h \delta n_n^h + \alpha_{\Lambda p}^h \delta n_p^h 
+ \alpha_{\Lambda \Lambda}^h \delta n_{\Lambda}^h \right) -   
\left(\alpha_{\Lambda n}^K \delta n_n^K + \alpha_{\Lambda p}^K \delta n_p^K 
+ \alpha_{\Lambda \Lambda}^K \delta n_{\Lambda}^K +    
\alpha_{\Lambda K^-} \delta n_{K^-} \right) &=& 0~.
\end{eqnarray} 
We express $\delta n_n^h$, $\delta n_p^h$, $\delta n_{\Lambda}^h$, 
$\delta n_p^K$, $\delta n_{\Lambda}^K$ and $\delta n_{K^-}$ in terms of
$\delta n_n^K$ using above six constraints. For this purpose, we solve 
a 6 $\times$ 6 matrix constructed out of above six relations and obtain
$\frac{\delta \mu}{\delta {n_n^K}}$. Similarly, we obtain 
$\frac{\delta \mu}{\delta {n_n^h}}$ in the mixed phase from the above 
constraints. This completes the calculation of relaxation time and bulk 
viscosity in the mixed phase. 

Next we calculate critical angular velocity as a function temperature and mass
of a rotating neutron star. The bulk viscosity damping timescale ($\tau_B$) 
due to the non-leptonic process involving $\Lambda$ hyperons and the bulk
viscosity profile as a function of $r$ are obtained following the 
Ref. \citep{Lin99,Lin02,Nar,DR1}. Further we take into account time scales 
associated with gravitational radiation ($\tau_{GR}$) \citep{Lin98} 
, bulk viscosity due to modified Urca process ($\tau_U$) involving only 
nucleons \citep{Saw,Kok} and the shear viscosity ($\tau_{SV}$) 
\citep{Lin98,Kok,And06} and define the overall r-mode time scale ($\tau_r$) as
\begin{equation}
{\frac {1}{\tau_r}} =  - {\frac {1}{\tau_{GR}}} + {\frac {1}{\tau_B}} + 
{\frac {1}{\tau_U}} + {\frac {1}{\tau_{SV}}}~. 
\end{equation}
Finally, solving $\frac {1}{\tau_r}$ = 0, we calculate the critical 
angular velocity above which the r-mode is unstable whereas it is stable below 
the critical angular velocity. 
\section{Results and Discussion}
For this calculation, nucleon-meson coupling constants are taken from 
Ref.\citep{Gle91} and this set is known as GM. Nucleon-meson coupling 
constants are determined by reproducing nuclear matter saturation
properties such as binding energy $E/B=-16.3$ MeV, baryon density $n_0=0.153$ 
fm$^{-3}$, asymmetry energy coefficient $a_{\rm asy}=32.5$ MeV, 
incompressibility $K=300$ MeV and effective nucleon mass $m^*_N/m_N = 0.70$. 
Further we need to know kaon-meson coupling constants and determine them using
the quark model and isospin counting rule. The vector coupling constants are
given by
\begin{equation}
g_{\omega K} = \frac{1}{3} g_{\omega N} ~~~~~ {\rm and} ~~~~~
g_{\rho K} = g_{\rho N} ~.
\end{equation}
The scalar coupling constant is obtained from the real part of
$K^-$ optical potential depth at normal nuclear matter density
\begin{equation}
U_{\bar K} \left(n_0\right) = - g_{\sigma K}\sigma - g_{\omega K}\omega_0 ~.
\end{equation}
Antikaons experience an attractive potential whereas kaons has a repulsive 
interaction in nuclear matter \citep{Fri94,Koc,Bat97,Waa,Li1,Li2,Fri99,Pal2}. 
The analysis of $K^-$ atomic data using a hybrid model \citep{Fri99} which 
combines the
relativistic mean field approach in the nuclear interior and a phenomenological
potential at low density, yielded the real part of the antikaon potential as
large as $U_{\bar K}(n_0) = -180 \pm 20$ MeV at normal nuclear matter density.
It was predicted that $K^-$ condensation might occur in neutron star matter for
strongly attractive antikaon potential $\sim -100$ MeV or more. 
In this calculation, we adopt the value of antikaon optical potential depth 
at normal nuclear matter density as $U_{\bar K}(n_0) = -160$ MeV. 
We obtain kaon-scalar meson coupling constant $g_{\sigma K} = 2.9937$ 
corresponding to this antikaon optical potential depth. 

On the other hand, hyperon-vector meson coupling constants are determined using
SU(6) symmetry of the quark model \citep{Dov,Sch94,Sch} and the scalar $\sigma$
meson coupling to $\Lambda$ hyperons is calculated from the hyperon 
potential depth in normal nuclear matter $U_{\Lambda}^N (n_0)$ = $-30$ MeV
obtained from hypernuclei data \citep{Dov,Chr}. The hyperon-$\sigma^*$ coupling
constant is determined from double $\Lambda$ hypernuclei data 
\citep{Sch93,Sch}.

The strange meson fields also couple with (anti)kaons.
The $\sigma^*$-K coupling constant determined from the decay of $f_0$(925) 
is $g_{\sigma^*K}=2.65$ 
and the vector $\phi$ meson
coupling with (anti)kaons $\sqrt{2} g_{\phi K} = 6.04$ follows from
the SU(3) relation \citep{Sch}. 

The onsets of $K^-$ condensate and $\Lambda$ hyperons in neutron star matter
are sensitive to the composition of matter and the strength of antikaon optical
potential depth \citep{Bani2}. It was further noted that the early appearance of
either $\Lambda$ hyperons or the $K^-$ condensate delayed the onset of the 
other to higher densities. In this calculation, for 
$U_{\bar K}(n_0) = - 160$ MeV, $\Lambda$ hyperons appear just after the onset of
$K^-$ condensation. The $K^-$ condensation sets in at a density 2.23 $n_0$ and
the mixed phase ends at 4.1 $n_0$. On the other hand, $\Lambda$ hyperons appear
at a density 2.51 $n_0$. It is worth mentioning here that we obtained 
qualitatively similar results with the GM parameter set corresponding to 
$K = 240$ MeV and
$U_{\bar K} (n_0) = -140$ MeV. However, this led to a soft EoS resulting in a 
neutron star mass below the accurately measured mass \citep{DR3}. 
The composition of $\beta$-equilibrated and charge neutral 
matter in the presence of $K^-$ condensate for $U_{\bar K}(n_0) = -160$ MeV 
is displayed in Figure 1. Negatively charged particles such as electrons and 
muons are depleted from the system with the onset of $K^-$ condensation 
and its rapid growth there after. At this stage, the proton density 
becomes equal to the density of $K^-$ mesons in the condensate. It is evident 
from the figure that $\Lambda$ hyperons populate the system just after the 
onset of $K^-$ condensation.    

The relaxation time for the non-leptonic process 
$n + p \rightleftharpoons p + \Lambda~$ is plotted with normalised baryon 
density in Figure 2 for $U_{\bar K}(n_0) = -160$ MeV and temperature 
$T = 10^{10}$ K. Here equations of state 
enter as inputs in the calculation of relaxation time in different phases. In 
particular, partial derivatives of pressure and chemical potentials with 
respect to neutron number density are calculated using EoS according to the 
prescription as discussed in section II. This figure shows relaxation time in
different phases i.e.  the pure antikaon condensed phase (light solid line) and 
the hadronic (bold solid line) and antikaon condensed parts of the
mixed phase (dashed line). We find that the values of relaxation time in the 
pure and mixed antikaon condensed phases are significantly smaller than that of 
the hadronic phase involving non-superfluid $\Lambda$ hyperons 
\citep{Nar,DR1}. The relaxation time for the non-leptonic weak process 
involving hyperons is inversely proportional to $T^2$ as given by Eq.(15). 

Figure 3 exhibits the hyperon bulk viscosity coefficient as a
function of normalised baryon density at a temperature $T = 10^{10}$ K. Similar
to Fig. 2, the bulk viscosity in the pure antikaon condensed phase 
(light solid line) and the hadronic (bold solid line) and antikaon condensed 
parts of the mixed phase (dashed line) are shown here. One can immediately see 
that the hyperon bulk viscosity in the antikaon condensed matter, irrespective 
of whether it is in the
pure or mixed phase, is suppressed compared with that of hadronic phase. This
suppression may be attributed to the superconducting phase i.e the $K^-$ 
condensed phase. The role of superfluidity on hyperon bulk viscosity was 
studied at length by several groups \citep{Han3,Nar,And06}. They also obtained
significant suppression in the hyperon bulk viscosity because one or more
superfluid particles were participating in non-leptonic weak processes 
involving hyperons. In our calculation, $K^-$ mesons in the Bose-Einstein 
condensed state is not a member of the non-leptonic weak process  
$n + p \rightleftharpoons p + \Lambda~$. Therefore, the suppression of hyperon
bulk viscosity in our case originates from the EoS in the $K^-$ condensed phase
which enters into the calculation of the chemical imbalance as given by 
Eq.(17). We further add that the factor $\omega \tau$ in 
the bulk viscosity coefficient given by Eq.(15) is negligible compared with 
unity over the whole range of baryon densities considered here. This leads to
a $1/T^2$ temperature dependence of the hyperon bulk viscosity. However, the
inversion of the temperature dependence was observed in some calculations
\citep{Han3,Nar} when the factor $\omega \tau$ is much greater than unity.
We find a jump in the bulk viscosity coefficient at the upper phase 
boundary of the mixed phase and pure $K^-$ condensed phase. This is attributed 
to kinks in the EoS and discontinuities in ($\gamma_{\infty} - \gamma_0$).
 
Now we discuss the results of damping time scale due to the hyperon bulk 
viscosity 
and critical angular velocity. This calculation needs the knowledge of   
the energy density profile, hyperon bulk viscosity profile and the structure 
of the rotating neutron star in question. For this purpose, 
we consider a neutron star of gravitational mass 1.60$M_{\odot}$ having baryon 
rest mass 1.76$M_{\odot}$ and central baryon density 3.50 $n_0$ and rotating at 
an angular velocity $\Omega_{rot} = 2652 s^{-1}$ from the sequence of rotating 
neutron stars calculated by the model of Stergioulas \citep{Ster95}. 
This neutron star contains both $\Lambda$ hyperons and $K^-$ condensate in its 
core as its central baryon density is well above the threshold densities of 
$K^-$ condensation and $\Lambda$ hyperons. The hyperon bulk viscosity profile 
of this neutron star as a function of equatorial distance for $T = 10^{10}$ K 
is displayed in Figure 4. The hyperon bulk viscosity in the hadronic and 
antikaon condensed parts of the mixed phase are shown by bold solid 
and dashed lines, respectively. 
Further we note that the bulk viscosity profile drops to zero value beyond 3.5 km
because the baryon density beyond this distance decreases below the threshold 
density of $\Lambda$ hyperons.

As soon as we know the energy density and bulk viscosity 
profiles, we obtain the damping time scale corresponding to the hyperon bulk 
viscosity and critical angular velocities as a function of temperature solving 
$1/\tau_r = 0$ for a rotating neutron star mass 1.60$M_{\odot}$.
Besides the hyperon bulk viscosity, we also consider the 
bulk viscosity due to modified Urca process involving nucleons as well as the 
shear viscosity to the total r-mode time scale as given by Eq.(34). 
The bulk viscosity due to modified Urca process plays an important 
role to damp the r-mode at higher temperatures. However, this process can not
suppress the r-mode instability below $10^{10}$ K because the corresponding 
damping time scale is longer than the 
gravitational radiation growth time scale. On the other hand, the shear 
viscosity coefficient is proportional to $T^{-2}$ and the damping time scale
$\tau_{SV}$ is also larger than $\tau_{GR}$ in the temperature range considered
here. The bulk viscosity damping time scale ($\tau_B$) due to the non-leptonic 
process $n + p \rightleftharpoons p + \Lambda~$ and $\tau_{GR}$ are comparable
at T $\sim 4 \times 10^{9}$ and below. Consequently, the r-mode instability is 
damped in this temperature regime by the hyperon bulk viscosity in $K^-$ 
condensed matter. Though the hyperon bulk viscosity is suppressed in the 
antikaon condensed phase, it is still a very efficient process to damp the 
r-mode instability.  

\section{Summary and Conclusions}       
We have investigated the hyperon bulk viscosity due to the non-leptonic weak 
process $n + p \rightleftharpoons p + \Lambda~$ in a $K^-$ condensed phase and
later applied it to study the r-mode instability in neutron stars. 
For the parameter set adopted here and antikaon optical potential depth
$U_{\bar K}(n_0) = -160$ MeV, $K^-$ condensation occurs before $\Lambda$ 
hyperons are populated in the system. We find that the hyperon bulk
viscosity coefficient in $K^-$ condensed matter is significantly suppressed 
compared with the non-superfluid hyperon bulk viscosity 
coefficient in the hadronic phase. Further we note that the hyperon bulk 
viscosity in the superconducting phase is still an efficient
process to damp the r-mode instability. 

\acknowledgments

DB thanks the Alexander von Humboldt Foundation for the support. He also
thanks Horst St\"ocker and Walter Greiner for their support and   
acknowledges the warm hospitality at the Institute for Theoretical Physics, 
J.W. Goethe University, Frankfurt am Main  and Frankfurt Institute for 
Advanced Studies (FIAS) where a part of this work was completed.

\clearpage

\begin{figure}
\epsscale{.80}
\plotone{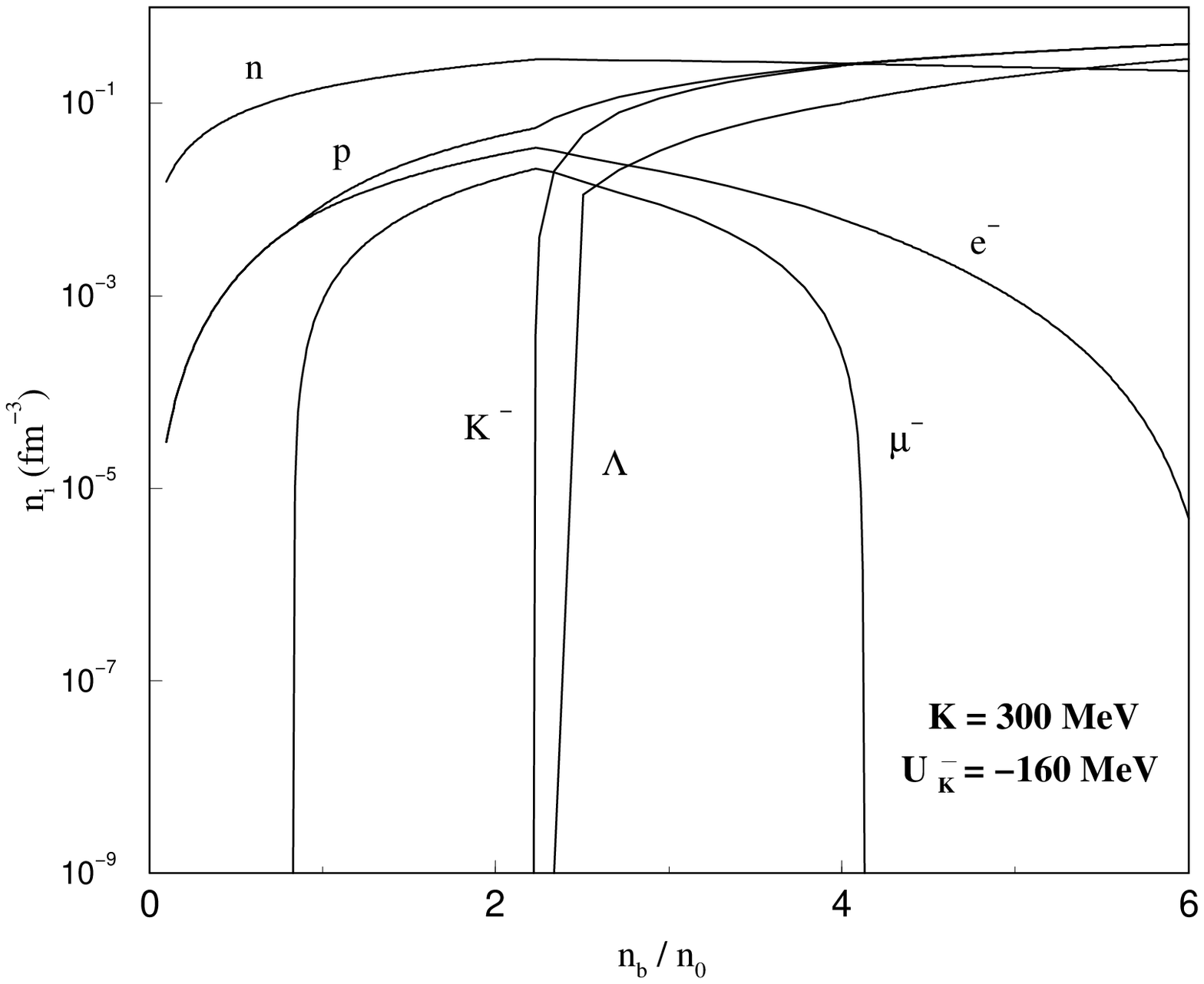}
\caption{Particles abundances are plotted with normalised baryon density for 
antikaon optical potential depth at normal nuclear matter density 
$U_{\bar K}(n_0) = -160$ MeV.}
\end{figure}

\clearpage

\begin{figure}
\epsscale{.80}
\plotone{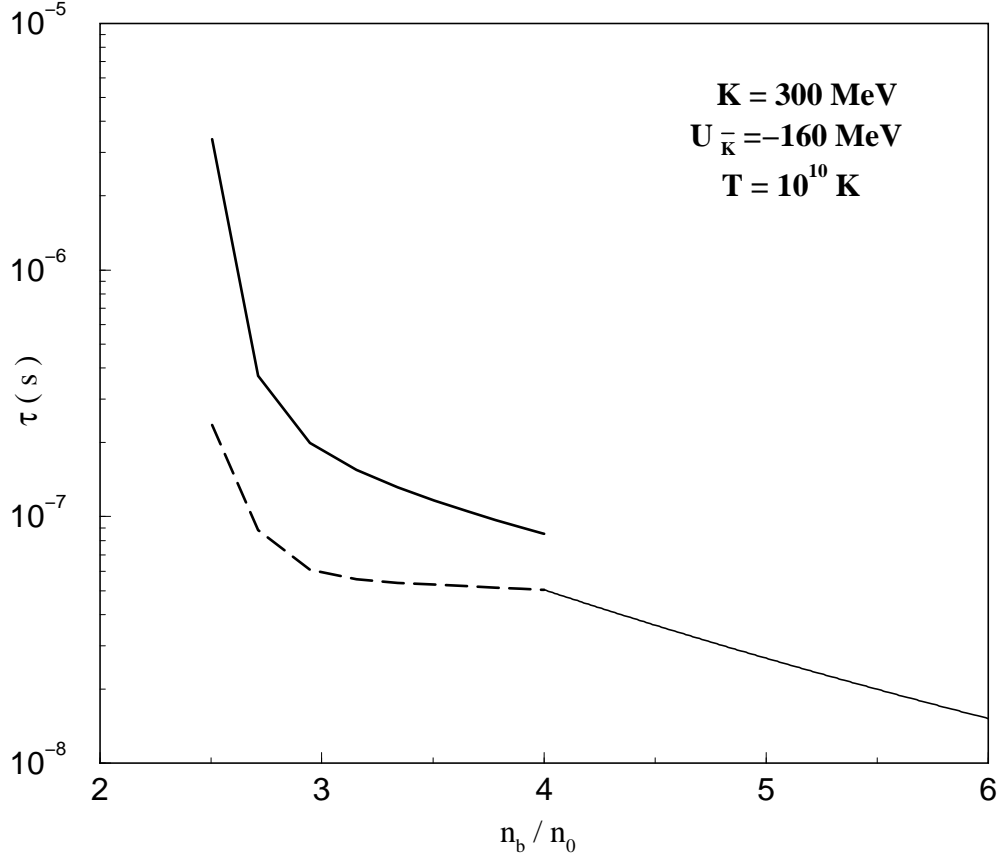}
\caption{Relaxation time is plotted with normalised baryon density for the 
non-leptonic process $n + p \rightleftharpoons p + \Lambda~$ and  
antikaon optical potential depth at normal nuclear matter density 
$U_{\bar K}(n_0)= -160$ MeV. The contributions of the pure antikaon condensed 
phase and the hadronic and antikaon condensed parts of the mixed phase
are shown by light solid, bold solid and dashed lines, respectively.}
\end{figure}

\clearpage

\begin{figure}
\epsscale{.80}
\plotone{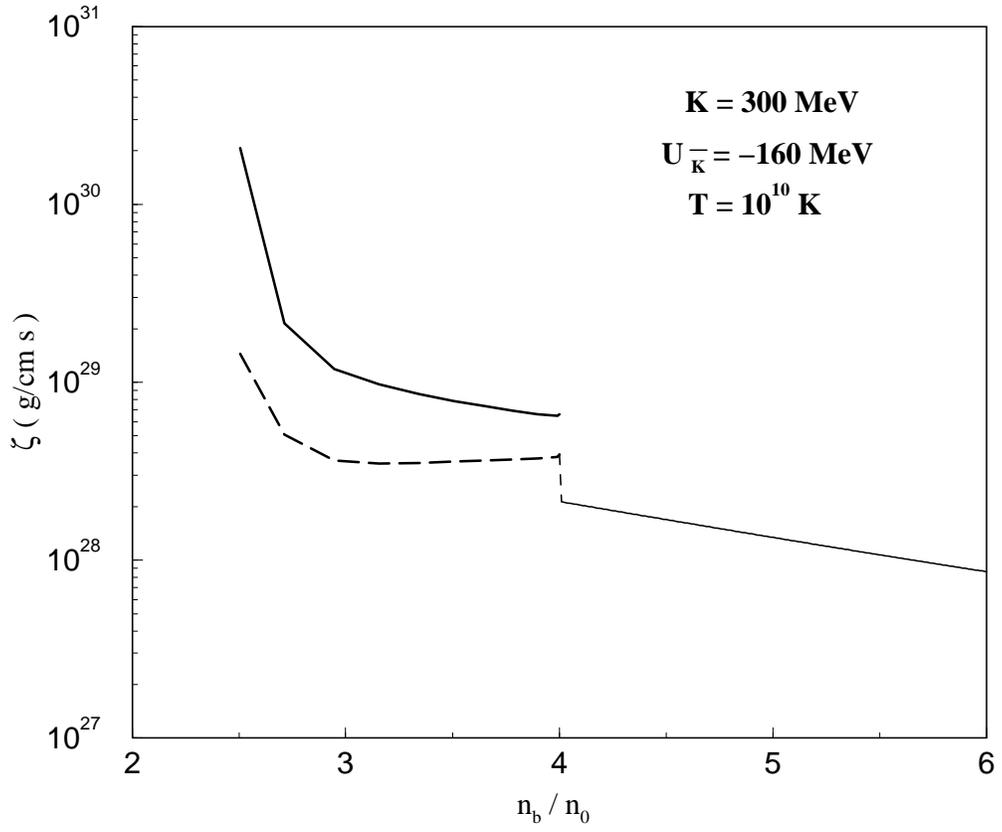}
\caption{Hyperon bulk viscosity coefficient is exhibited as a function of 
normalised baryon density at a temperature $10^{10}$ K and
antikaon optical potential depth at normal nuclear matter density 
$U_{\bar K}(n_0) = -160$ MeV. Different lines have the same meaning as in 
Fig. 2.} 
\end{figure}

\clearpage

\begin{figure}
\epsscale{.80}
\plotone{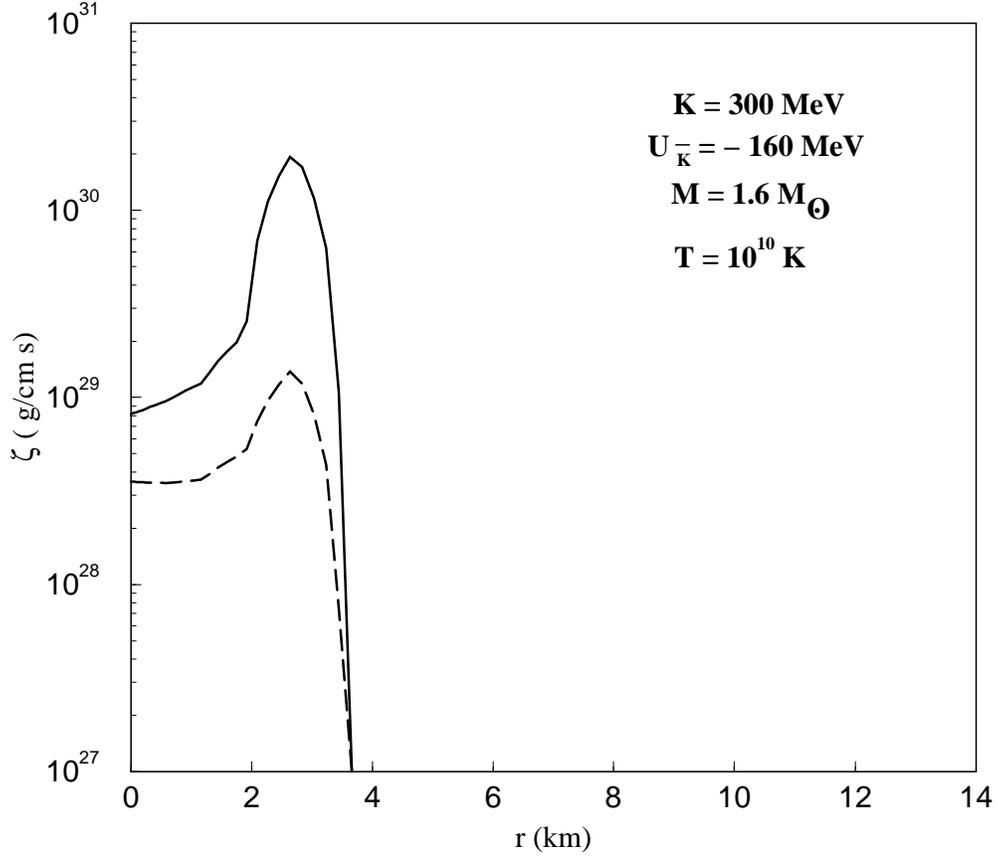}
\caption{Hyperon bulk viscosity profile is shown with equatorial distance for
a rotating neutron star of mass 1.60 M$_{\odot}$ at a temperature $10^{10}$ K 
and antikaon optical potential depth $U_{\bar K}(n_0) = -160$ MeV. The 
hadronic and antikaon condensed parts of the mixed phase 
are shown by bold solid and dashed lines, respectively.}
\end{figure}

\clearpage

\begin{figure}
\epsscale{.80}
\plotone{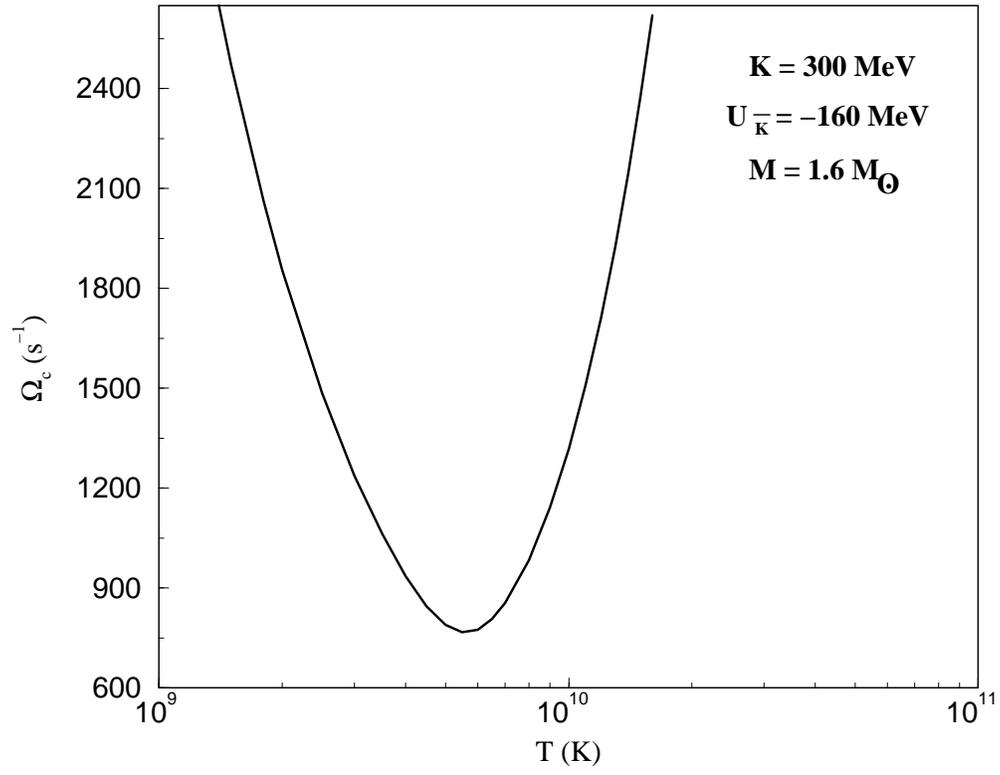}
\caption{Critical angular velocity for 1.60 M$_{\odot}$ neutron star is 
plotted as a function of temperature.}
\end{figure}

\end{document}